# Determination of the Surface Plasmons Polaritons extraction efficiency from a self-assembled plasmonic crystal

Hugo Frederich[1,2], Fangfang Wen[1,2], Julien Laverdant[1,2,3], Willy Daney de Marcillac[1,2], Catherine Schwob[1,2], Laurent Coolen[1,2], and Agnès Maître[1,2,*]

[1]*Université Pierre et Marie Curie-Paris 6, UMR 7588, INSP, 4 place Jussieu, PARIS cedex 05, France*
[2]*CNRS, UMR 7588, INSP, Paris cedex 05, France*
[3]*Institut Lumière Matière, UMR5306 Université Lyon 1-CNRS,Université de Lyon, 69622 Villeurbanne cedex, France*

*Corresponding author : agnes.maitre@insp.jussieu.fr

**Abstract**. We experimentally measure and analytically describe the fluorescence enhancement obtained by depositing CdSe/CdS nanocrystals onto a gold plasmonic crystal, a two-dimensional grating of macroscopic size obtained by gold deposition on a self-assembled opal. We show evidences of nanocrystals near-field coupling to the gold Surface Plasmons Polaritons (SPP) followed by grating-induced SPP re-emission to far-field. We develop a theoretical framework and an original method in order to evaluate, from photoluminescence experiments, the SPP extraction efficiency of a grating.

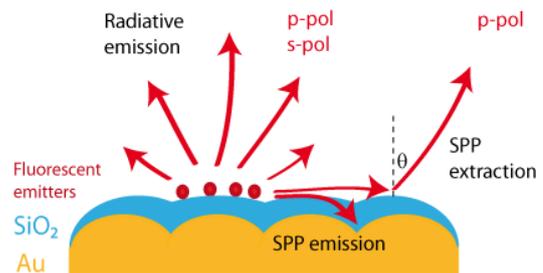

## 1. Introduction

Surface Plasmons Polaritons (SPP)[1] are widely used to enhance light-matter interactions in various domains such as bio-imaging,[2] photovoltaics,[3] photocatalysis,[4] surface enhanced Raman spectroscopy,[5] single photon sources,[6, 7] and Light Emitting Diodes[8]. The latter examples are part of the rapidly growing field of SPP-assisted light emission.[9]

When an emitter is close to a metallic film, its spontaneous emission dynamics is considerably accelerated. However, a large part of the emission energy is then injected into the SPP modes. An important challenge is then to extract the energy from the evanescent SPP modes in order to radiate it into far field.[10]

Similarly, in the case of dielectric waveguides, light extraction is also an important issue since luminescent materials usually have high optical indices and guide light. Periodic structures offer the possibility to extract guided electromagnetic modes and to redirect them to far field.[11]

In order to couple the SPP modes to far-field light, periodic structures (plasmonic crystals) can be used.[12] Many particular structures have been experimented in the SPP extraction context: from 1D groove gratings,[13] or nanohole arrays,[14] to dielectric gratings on top of planar metallic surfaces[15] or hybrid plasmonic-photonic structures.[16,17]

For many applications, the spectral and angular broadness of SPP coupling and the ability of fabricating large active surface areas are essential. Among other techniques, self-assembly offers interesting prospects to obtain complex structures allowing broadband SPP re-emission.[18-24]

Very few authors have proposed a quantitative evaluation [25-27] of the SPP extraction efficiency from periodic structures. Nevertheless, this quantity is important to evaluate and characterize the light emission enhancement.[10]

Most studies quantifying SPP-assisted light emission take place in the context of Light Emission Diodes.[8, 12, 28, 29] The measured quantity is usually not the extraction efficiency but the overall yield of the device.

Yet, the SPP extraction efficiency has been specifically measured experimentally in two studies [25, 26] involving one- and two-dimensional gratings elaborated on silver thin films deposited on prisms. The method applies to optically thin gratings: it consists of exciting SPP in the Kretschmann configuration on one side of the silver film and measuring the grating out-coupling intensity on the other side. Extraction efficiencies up to 60 % have been found for monochromatic light at specific angles.

To our knowledge, no other experimental method has been proposed to determine SPP extraction efficiency in more general cases.

In this paper, we study experimentally light-SPP coupling on a plasmonic crystal obtained from a self-assembly technique. And we propose an original experimental method to determine the SPP extraction efficiency of plasmonic crystals using nano-emitters as probes.

We first describe the fabrication of the sample and we present its optical properties. Next, we study the angle- and polarization-dependent fluorescence emission of semiconductor nanocrystals deposited on the plasmonic crystal. Then, we develop an analytical model of the emission intensity. Finally, we describe the experimental method to evaluate the extraction efficiency and apply it to our sample.

**2. Plasmonic crystal fabrication and characterization**

The principle of the plasmonic crystal fabrication is to evaporate an optically thick layer of gold onto a self-assembled periodical array of silica spheres (artificial opal) which is used as a template to impose a periodic corrugation to the gold layer. Preliminary, a first 100 nm-thick smoothing layer of silica is deposited on the artificial opal. It is used to reduce the groove depth of the grating.

A 150 nm-thick layer of gold is deposited afterwards. As its thickness is larger than the SPP skin depth, it is considered optically infinite. The two-dimensional triangular grating which is obtained is presented on **figure 1**.

The nano-structured surface is about 1 cm² and the mean size of the ordered domains is approximately 20 μm. At the millimetre scale, the plasmonic crystal is then poly-crystalline and fully isotropic in the azimuthal direction.[18] The grating period and the mean groove depth of the plasmonic surface are evaluated to $390 \pm 10$ nm and $75 \pm 10$ nm by scanning electron and atomic force microscopies, respectively.

Finally, a third 70 nm-thick layer of silica is deposited on the gold to act as a spacer.

The optical properties of such gold surfaces (without spacers) have been studied specifically in a previous study showing that the grating couples far-field light to SPP modes in p-polarization, consistently with an analytical model.[18]

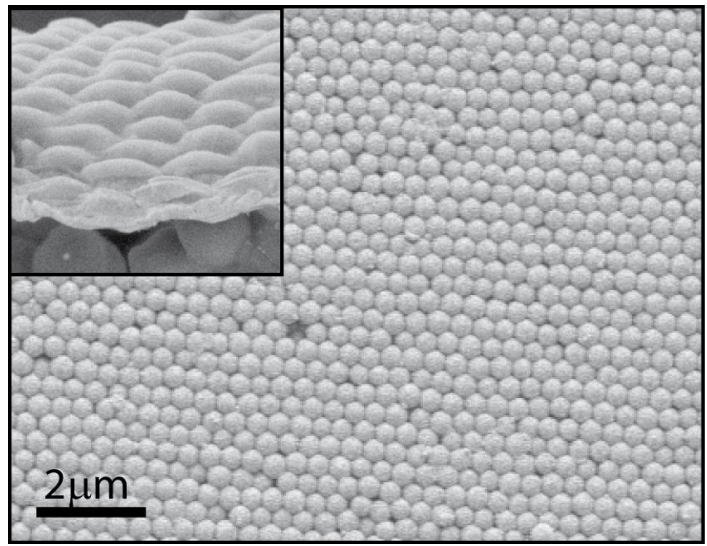

**Figure 1.** SEM micrographs of the plasmonic crystal surface.

Preliminary experiments, such as polarization- and angle-resolved specular reflectivity, are performed to characterize the plasmonic properties of the surface. The spectra acquired for incidence angles θ varying from 20° to 70° are shown in **figure 2**, they are isotropic in the azimuthal direction.

The faint and slowly shifting absorption dip seen on the s-polarized measurement below 600 nm was attributed, in Ref. [18], to Localized Surface Plasmon (LSP) modes.

In p-polarization, a broad absorption dip is observed; its width at a given angle spans from 30 nm to 50 nm. This coupling broadness is an outcome of the use of a self-assembled substrate for the plasmonic crystal.

This dip is red-shifted when θ increases. Its central wavelength is represented as a function of $\sin\theta$ in the inset of figure 2.b (circles) along with the calculated SPP coupling wavelength for a 2D plasmonic crystal covered with 70 nm of silica (solid line). The model used for the calculation consists in applying the grating phase-matching conditions to SPP's dispersion relation for a flat surface. It is valid in the low corrugation regime (reached when the mean groove depth is lower than the SPP skin depth[18]). This regime is achieved when surface features can scatter SPP modes with little modification of their dispersion relation.

The good agreement, without any fitting parameter, between calculation and experiment (less than 2.5% relative difference above 600 nm) evidences that i) the dip is caused by the grating-induced coupling of the light to the SPP modes; ii) the plasmonic crystal is in the low corrugation regime above 600 nm.

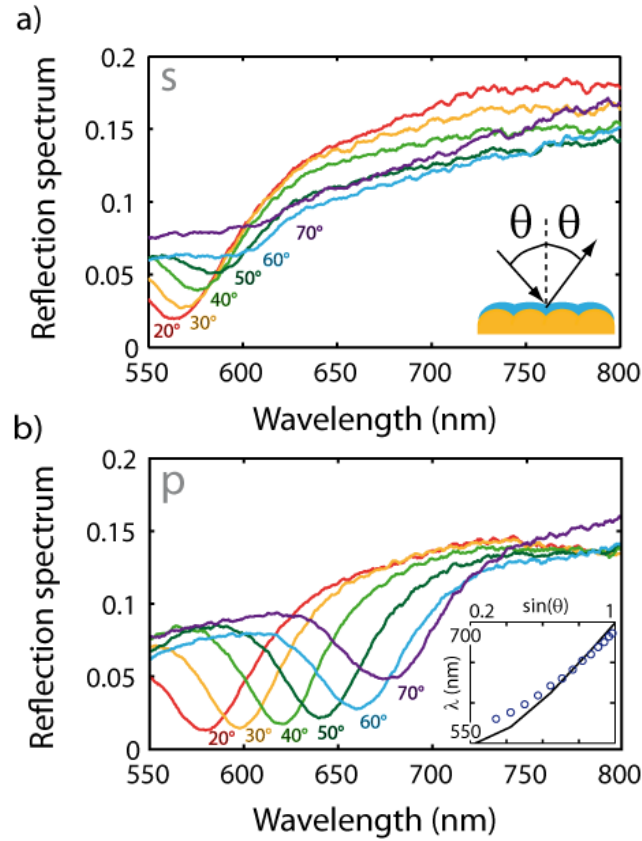

**Figure 2.** Polarized reflectivity spectra of the plasmonic sample at incidence angles θ ranging from 20° to 70°, with steps of 10°. a) In s-polarization. b) In p-polarization. Inset: experimental (circles) and calculated (solid line) coupling wavelengths as a function of $\sin\theta$.

## 3. Photoluminescence measurements

In this section, we describe the photoluminescence measurements performed on CdSe/CdS nanocrystals deposited onto the plasmonic crystal described previously.

We use semiconductor nanocrystals as fluorophores for their high brightness and their versatility.[30] The nanocrystals consist of a CdSe core and a CdS shell for a total diameter of about 10 nm.[31] They do not blink and they have a large absorption cross-section as well as a high quantum efficiency.[32] Nanocrystals are directly drop cast on the surface of the sample. The ligands on the surface of the nanocrystals prevent them from forming clusters. The concentration of the nanocrystals solution was chosen to avoid stacking and to obtain a layer with a thickness of approximately 10 nm.

Given the silica spacing layer thickness (70 nm), this preparation allows a near-field coupling of the nanocrystals fluorescence emission to the gold SPP modes and avoids any quenching effect. Nanocrystals with an emission spectrum spanning from 600 nm et 660 nm are chosen so that they are coupled to the SPP modes only (and not to the LSP modes).

A polarization- and angle-resolved photoluminescence (PL) measurement is carried out (see Inset of **figure 3**). The nanocrystals are excited at fixed incidence with a laser diode (10 µW at 405 nm) focused on the sample with a spot of approximately 1 mm². The emitted light is collected, over the range $\theta \in [-80°, 80°]$, through a polarizer and an optical fiber (N.A. 0.17) rotating at 15 cm from the sample, it is then analyzed with a spectrometer.

The polarized emission intensities $I_p(\lambda, \theta)$ and $I_s(\lambda, \theta)$ are presented in figure 3 for two particular angles: 0° and 50°. All the spectra are corrected by the polarization-dependent apparatus function of the set-up.

For $\theta = 0°$, both polarizations are equivalent, therefore $I_p$ and $I_s$ are superimposed.

For $\theta = 50°$, the p-polarized spectrum shape is modified: $I_p$ is less intense than $I_s$ below 635 nm and more intense above. As a result, the maximum of $I_p$ is redshifted of a few nanometers compared to the $\theta = 0°$ case.

In the next section, we show that these features can be interpreted as the effects of the gold surface structure.

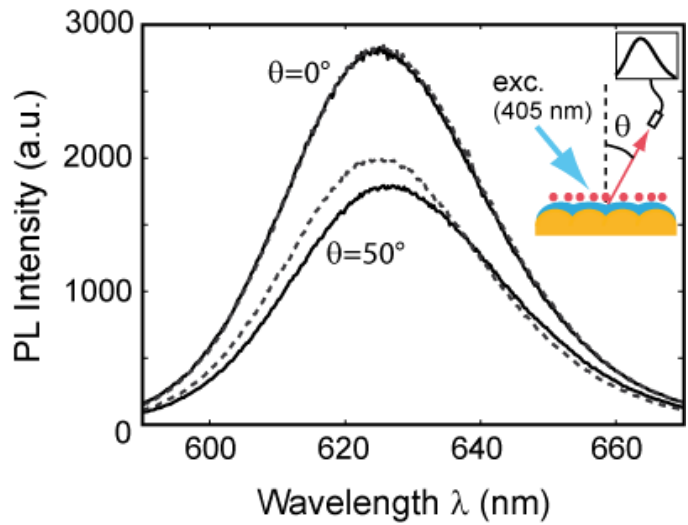

**Figure 3.** Polarized photoluminescence (PL) measurements. Emission spectra measured at θ = 0° and θ = 50° from the CdSe/CdS nanocrystals deposited on the gold plasmonic crystal for the two polarizations of light: $I_s(\lambda, \theta = 0°)$ and $I_s(\lambda, \theta = 50°)$ (dashed lines); $I_p(\lambda, \theta = 0°)$ and $I_s(\lambda, \theta = 50°)$ (solid lines). Inset: PL measurement scheme.

## 4. Evidence of SPP-assisted emission

The purpose of this section is to evidence SPP-assisted emission: a fraction of the SPP waves excited in near-field by the nanocrystals is radiated in far-field by the grating. This SPP re-emission is assumed to be p-polarized only since the s-polarized light is not coupled to SPP modes. Therefore, s-polarized measurements will be used as references.

The modification of the p-polarized spectrum acquired at $\theta = 50°$ (figure 3) is expected to come from this effect. To validate this assumption, the ratio between the p- and s-polarized spectra, $I_p/I_s$, is computed. **Figure 4** shows this ratio for different wavelengths as a function of $\theta$. The dashed line at 50° corresponds to the particular spectra plotted in figure 3 and confirms that $I_s(\lambda, \theta = 50°)$ is greater (respectively, smaller) than $I_p(\lambda, \theta = 50°)$ for $\lambda$ below (respectively above) 640 nm. The main feature of figure 4 is the angle- and wavelength-dependent peak of p-polarized emission surplus. We show in the following that this p-polarized emission surplus is associated to grating-induced SPP re-emission.

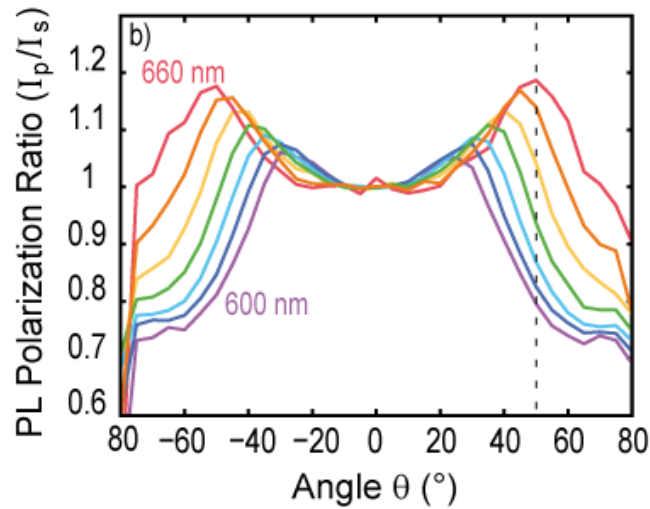

**Figure 4.** Emission polarization ratio as a function of the angle of observation θ for six different wavelengths (λ = 600 to 660 nm with 10 nm steps), obtained for a given angle by dividing the p-polarized spectrum by the s-polarized one. The dashed vertical line marks the 50° case represented in Figure 3.

**Figure 5** shows a comparison between the photoluminescence and the reflectivity measurements described in the previous section. The wavelengths obtained for the SPP coupling in the reflectivity spectra of figure 2.a are plotted as a function of $\sin\theta$ (red points) The latter curve can be analysed as a SPP dispersion relation. The wavelengths of the $I_p/I_s$ peak observed in figure 4 are also plotted on the same figure as a function of $\sin\theta$. The two measurements show a clear agreement. This proves that the $I_p/I_s$ peak observed in figure 4 is due to the p-polarized SPP re-emission by the grating.

Moreover, SPP modes are very sensitive to the optical index of the material present in the vicinity of the surface. Therefore, the slight redshift of the SPP dispersion relation obtained by fluorescence measurements (see figure 5) can be attributed to the high optical index of the nanocrystals added on top of the sample.

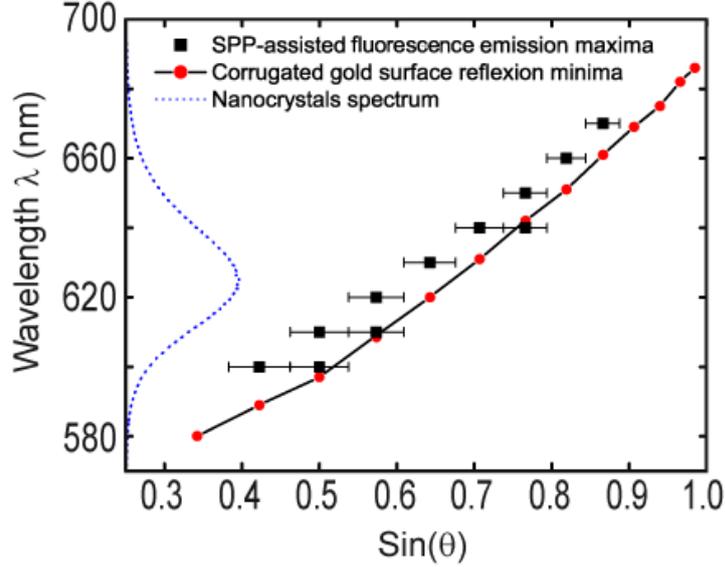

**Figure 5.** Wavelength of the SPP-photon coupling maximum as a function of the angle of observation in the cases of the reflectivity (dots) and the fluorescence (squares) measurements.

In this section, we have shown that among the SPP excited in near-field by the nanocrystals, a fraction is redirected to the far-field by the grating and contributes to the overall photoluminescence of the sample. The emission can then be decomposed into a direct part (the radiative emission of the nanocrystals), and an indirect part (the emission recovered from the SPP).

In the following sections, we develop a model to distinguish and then quantify those two parts.

**5. Emission model**

In order to distinguish and separate each contributions of light emission into $I_p$ and $I_s$, a theoretical framework is established in this section. The detailed expressions of $I_p$ and $I_s$ are given at the end of the section.

When considering nanocrystals as two-level emitting systems, we can define excitation and decay rates $r$ and $\Gamma$, respectively. Here, we consider $\Gamma = \Gamma_{\text{rad}} + \Gamma_{\text{SPP}}$ with $\Gamma_{\text{rad}}$ and $\Gamma_{\text{SPP}}$ defined as the partial decay rates in the radiative and the SPP channels, respectively. Without any change in the method that we propose, a non-radiative decay channel could be added in order to take into account emitters intrinsic non-radiative processes or coupling to Lossy Surface Waves (LSW). In the present case, we neglect these effects since the nanocrystals have high quantum efficiency[32] and the 70 nm-thick silica spacing layer prevents any LSW coupling.[9]

The overall photon emission rate can be obtained from the combination of the excitation and the radiative decay rates. In the stationary regime, the population dynamics of a two-level system yields a photon emission rate $r\Gamma_{\text{rad}}/(r + \Gamma)$.

The excitation rate $r$ is expressed as $r = P\sigma/\hbar\omega_{exc}$, where $P$ is the excitation power density, $\hbar\omega_{exc}$ is the excitation photon energy and $\sigma$ is the nanocrystal absorption cross-section. In the present study, $P = 10$ Wm$^{-2}$ and $\sigma$ is of the order of $10^{-19}$ m$^{-2}$ (see Ref. [33]), therefore $r$ is of the order of $10$ s$^{-1}$. As $\Gamma$ is of the order of $10^8$ s$^{-1}$, the system is in the low pumping regime (characterized by $r \ll \Gamma$) even in the case of a local enhancement of the excitation field. This implies that the photon emission rate can be approximated by $r\Gamma_{\text{rad}}/\Gamma$.

Correspondingly, the plasmon emission rate is approximated by $r\Gamma_{\text{SPP}}/\Gamma$.

The angular distributions of the radiative emission for both polarizations $\alpha_{rad,p}(\lambda,\theta)$ and $\alpha_{rad,s}(\lambda,\theta)$ are introduced. The integral of the distributions $\alpha_{rad,p}(\lambda,\theta)$ and $\alpha_{rad,s}(\lambda,\theta)$ over the entire space is set to 1. In the present case, the radiation develops into the top half-space only and the sample is isotropic in the azimuthal direction, hence:

$$2\pi \int_0^{\pi/2} \left(\alpha_{rad,p}(\lambda,\theta) + \alpha_{rad,s}(\lambda,\theta)\right) \sin\theta \, d\theta = 1 . \tag{1}$$

Contrarily to radiative channel, the energy decayed in the SPP channel is not fully recovered optically. It is mainly absorbed in the metal but a fraction of it is radiated by the grating towards the top half-space. This fraction, comprised between 0 and 1, is written $\tilde{\alpha}(\lambda)$ as :

$$\tilde{\alpha}(\lambda) = 2\pi \int_0^{\pi/2} \alpha_{SPP}(\lambda,\theta) \sin\theta \, d\theta , \tag{2}$$

where $\alpha_{SPP}(\lambda,\theta)$ is the angular distribution of the SPP re-emission.

The function $\tilde{\alpha}(\lambda)$ represents the SPP extraction efficiency, it is determined in the next section.

The function $\alpha_{SPP}(\lambda,\theta)$ is defined to describe the grating-induced plasmon re-emission which only occurs in p-polarization above 600 nm (see figure 2). As a consequence, we consider in this study that $\alpha_{SPP}(\lambda,\theta)$ plays a role in $I_p$ expression only.

Finally, the polarized fluorescence intensities $I_p$ and $I_s$ can be written as:

$$I_s(\lambda,\theta) = S(\lambda)\frac{r\Gamma_{rad}}{\Gamma}\alpha_{rad,s}(\lambda,\theta) \tag{3}$$

$$I_p(\lambda,\theta) = S(\lambda)\left\{\frac{r\Gamma_{rad}}{\Gamma}\alpha_{rad,p}(\lambda,\theta) + \frac{r\Gamma_{SPP}}{\Gamma}\alpha_{SPP}(\lambda,\theta)\right\}, \tag{4}$$

where $S(\lambda)$ is the spectral distribution of the emission intensity for the nanocrystals population embedded in a homogeneous medium.

$I_p$ expression clearly separates the emission into its direct part (the radiative emission), and its indirect part (the SPP re-emission).

In the next section we propose a method to isolate, from experimental photoluminescence results, the function $\alpha_{SPP}(\lambda, \theta)$ representing the grating-induced SPP re-emission. Then, we perform an experimental evaluation of the SPP extraction efficiency: $\tilde{\alpha}(\lambda)$.

**6. SPP Extraction efficiency determination**

Photoluminescence experiments and calculated emission diagrams are used to quantitatively evaluate the SPP extraction. To do so, a comparison is performed between the plasmonic crystal and a planar reference (a gold semi-space covered with a 70 nm-thick silica film).

The emission from the planar reference can be described by $I_s^{\text{ref}}$ and $I_p^{\text{ref}}$ defined as :

$$I_{s/p}^{\text{ref}}(\lambda, \theta) = S(\lambda) \frac{r\Gamma_{rad,s/p}^{\text{ref}}}{\Gamma} \alpha_{rad,s/p}^{\text{ref}}(\lambda, \theta) \ , \tag{5}$$

where the decay rates $\Gamma$, $\Gamma_{rad,s/p}$ can be calculated from Ref. [34] and the angular distributions of emission $\alpha_{rad,s/p}$ from Ref. [36]. The term $\alpha_{SPP}^{\text{ref}}$ is absent as SPP re-emission does not occur on the planar reference.

To perform the SPP extraction evaluation, we assume that : except the SPP re-emission phenomenon, photoluminescence is identical from the plasmonic crystal and from a planar reference. In detail, this assumption can be reduced to two equalities :

$$\alpha_{rad,s/p} = \alpha_{rad,s/p}^{\text{ref}} \tag{6}$$

and

$$\Gamma_{rad/SPP} = \Gamma_{rad/SPP}^{\text{ref}} \ . \tag{7}$$

The decay rates and the angular distributions of the radiative emission are identical for both types of surface. The effect of the corrugation is thus assumed to lie only in the $\alpha_{SPP}$ term, which is zero for the planar reference.

In this section, we first show results supporting the two parts of the hypothesis (equations (6) and (7)) for the plasmonic crystal studied here. Then, we use this model to evaluate the SPP extraction efficiency from our photoluminescence measurements and discuss the results.

To verify the validity of equation (6), the theoretical emission diagram (angular distribution of emission) of an isotropic assembly of emitters on a planar reference is calculated for the two polarizations.[34] These calculations take into account the interferences effects in the structure by including both interfaces (gold/silica and silica/air) in effective reflection coefficients.[35] They show perfect agreement with experimental emission diagrams measured on planar structures (not shown here).

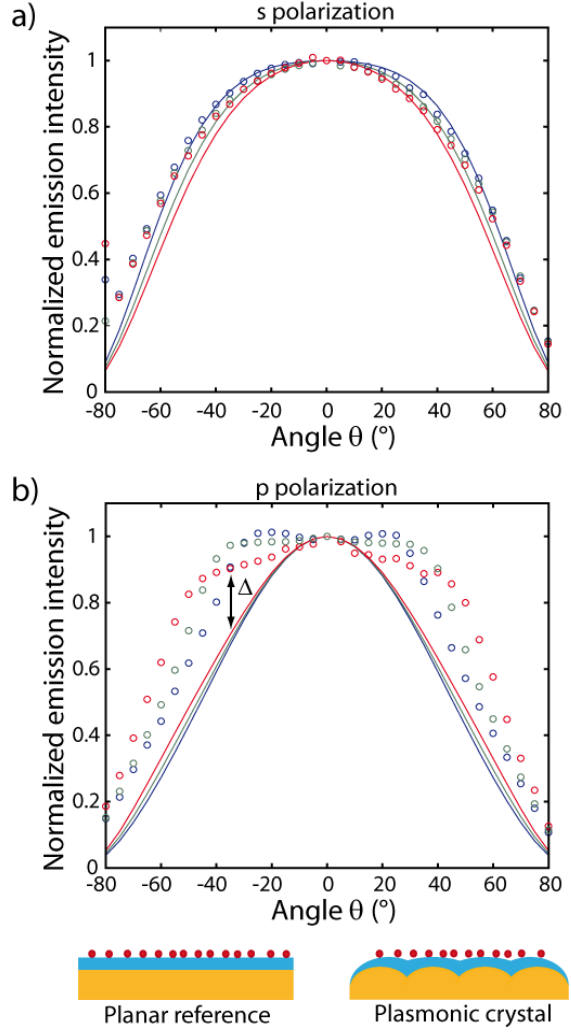

**Figure 6.** Comparison between emission diagrams on a planar reference (calculated, solid lines) and on the plasmonic crystal (experimental, circles) at 600 nm (blue), 630 nm (green), and 660 nm (red) and in a) s-polarization and b) p-polarization. Normalized emission diagrams are plotted here: $I_{s/p}(\lambda, \theta)/I_{s/p}(\lambda, 0)$. The difference $\Delta$ between the reference and the plasmonic crystal is due to SPP re-emission.

**Figure 6** presents the calculated normalized emission diagrams for λ = 600, 630, and 660 nm (solid lines). It is important to note that normalized quantities are used here to screen possible plasmonic enhancement of excitation $r$, so that only the emission enhancement is probed.

In s-polarization, the emission is only composed of the direct part, therefore the intensities measured from the nanocrystals on the plasmonic crystal (circles) are in very good agreement with the calculations for a planar reference. We deduce that low corrugation does not influence the direct radiative emission distribution, which validates the hypothesis of equation (6) in s-polarization.

We assume that the hypothesis remains valid for p-polarization, although it cannot be checked as the SPP re-emission component adds up to the radiative emission.

To validate the hypothesis of equation (7), the decay of the photoluminescence (PL) is measured in three cases: i) nanocrystals in solution (hexane), ii) nanocrystals deposited on the plasmonic crystal, and iii) nanocrystals deposited on a planar reference. The excitation light is delivered by a nitrogen laser emitting 1 ns pulses at 337 nm. The emission is collected and decomposed by a spectrometer to select the light emitted at 625 nm which corresponds to the center of the emission spectrum. The time-resolved intensity acquisition is achieved with a photo-multiplier module. The experimental results are shown in **figure 7**. The PL decays are not mono-exponential due to the heterogeneity of the nanocrystals and to their random orientations on gold.[7] These results reveal an approximately 2-fold acceleration of the emission when comparing the nanocrystals in solution and deposited close to the gold interface. Moreover, the PL decay on the planar reference and on the plasmonic crystal are similar. This experiment evidences the near-field coupling of the nanocrystals fluorescence to the gold surface modes in both cases.

In our case, surface modes reduce to SPP since LSP do not exist at the considered wavelength (see figure 2) and the 70 nm-thick spacer prevents any coupling between the nanocrystals and the LSW.[9] Hence, the dynamic of the emission is described by the rate $\Gamma = \Gamma_{rad} + \Gamma_{SPP}$. Let us discuss these three terms separately. First, the experiments show that $\Gamma$ takes similar values on the plasmonic crystal and on the planar reference. Then, given the fact that the SPP modes are evanescent, $\Gamma_{SPP}$ is mainly determined by the distance between the emitters and the gold surface. This distance is fixed by the silica layer, hence $\Gamma_{SPP}$ is supposed to be equal in both cases. Finally, this implies that $\Gamma_{rad}$ is equivalent on the two surface studied and validates equation (7).

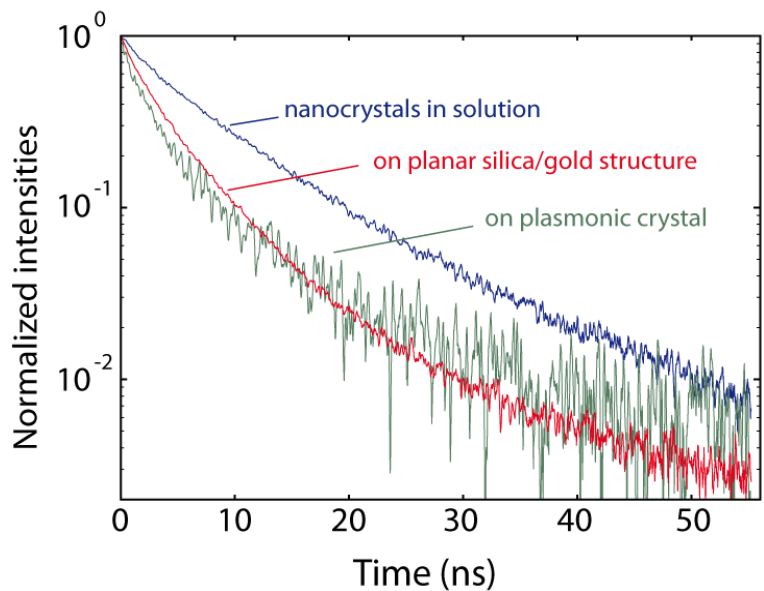

**Figure 7.** Photoluminescence decay measurements at 625 nm performed on the nanocrystals in hexane, on a planar silica/gold structure, and on the plasmonic crystal.

Two independent results support the hypothesis materialised by equations (6) and (7). We now use our model to evaluate SPP extraction.

The quantitative evaluation of the SPP re-emission process is achieved from the data of figure 6.b. It relies on the fact that Δ, the difference between the normalized emission diagram in p-polarisation of the sample and the one of the reference, is attributed to the SPP re-emission. Since $\alpha_{SPP}(\lambda, 0) = 0$ for λ = 600 nm, Δ is expressed as :

$$\Delta = \frac{I_p(\lambda,\theta) - I_p^{\text{ref}}(\lambda,\theta)}{I_p(\lambda,0)} \quad . \tag{8}$$

From the expressions of $I_p$ and $I_p^{\text{ref}}$ (equations (4) to (7)), we determine:

$$\Delta = \frac{\Gamma_{\text{SPP}}}{\Gamma_{\text{rad}}} \frac{\alpha_{SPP}(\lambda,\theta)}{\alpha_{rad,p}(\lambda,0)} \quad . \tag{9}$$

**Figure 8.a** represents this quantity as a function of the angle of observation θ for λ = 600, 630 and 660 nm. The wavelength- and angle-dependent peak of p-polarized emission associated to SPP is clearly visible.

It is then possible to extract the function $\alpha_{SPP}(\lambda, \theta)$ from Δ, in the context of hypothesis (6) and (7), by calculating the terms $\Gamma_{\text{SPP}}/\Gamma_{\text{rad}}$ as in Ref. [36], and $\alpha_{rad,p}(\lambda, 0)$ as in Ref. [34].

The function $\alpha_{SPP}(\lambda, \theta)$ is plotted on figure 8.b for λ = 600 and 660 nm. The integration over θ is then performed to determine the SPP extraction efficiency $\tilde{\alpha}(\lambda)$. In our particular case, this method gives extraction efficiencies comprised between $\tilde{\alpha}(600 \text{ nm}) = 4.5\%$ and $\tilde{\alpha}(660 \text{ nm}) = 6\%$. Note that these values are reached on very broad spectral and angular ranges.

The order of magnitude measured for $\tilde{\alpha}$ is consistent with the analytical calculations of Khurgin and co-workers.[10] According to this work, the values of $\tilde{\alpha}$ can be higher for longer wavelengths since the propagative constant of the SPP modes, their confinement, and the losses are low. Nevertheless, low confinement implies that the emission acceleration is weaker. Hence, an optimum wavelength exists to maximise the total light emission surplus due to SPP re-emission structures.

From a practical perspective, a relevant figure of merit is thus the re-emission ratio $\tau(\lambda) = \frac{\Gamma_{SPP}}{\Gamma_{rad}} \tilde{\alpha}$. It expresses the intensity ratio of the SPP re-emission to the radiative emission. It corresponds to the total light surplus due to the grating introduction compared to the planar case. In our particular example, we found $\tau(600 \text{ nm}) \approx 15\%$ and $\tau(660 \text{ nm}) \approx 20\%$.

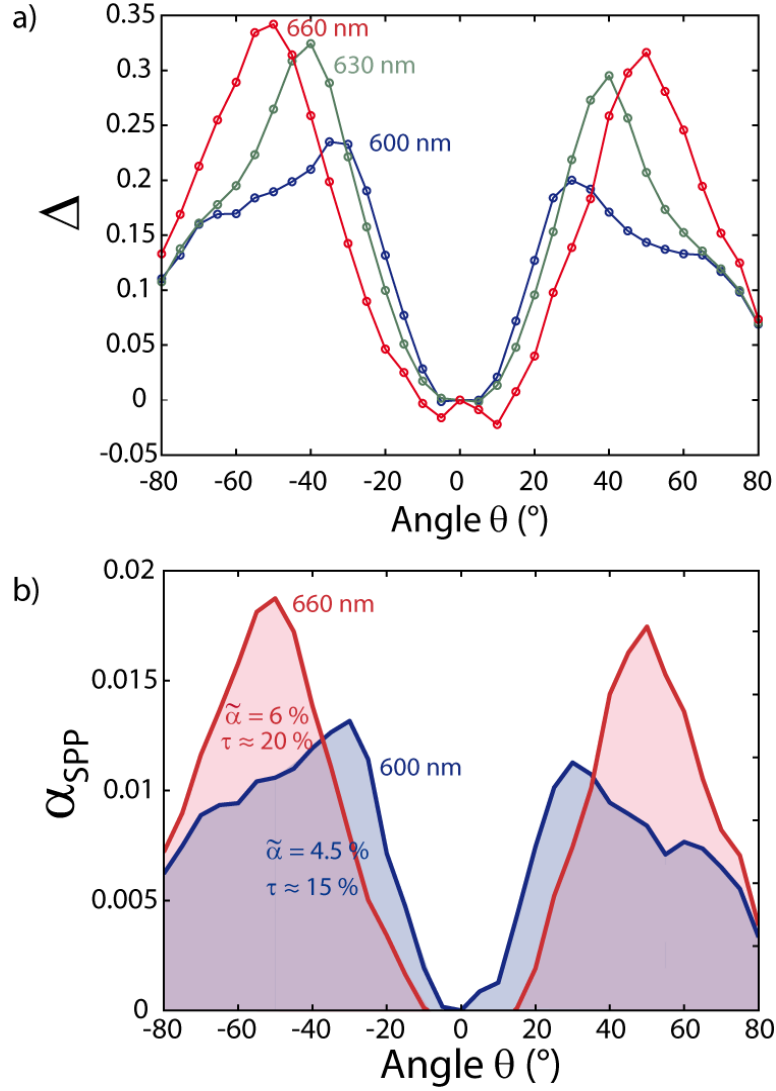

**Figure 8.** a) Quantity Δ represented as a function of the angle of observation θ for λ = 600, 630 and 660 nm. Δ is obtained by considering the difference between the normalized emission diagram of the plasmonic crystal and the one of the reference (figure 6.b). b) Representation of $\alpha_{SPP}(\lambda,\theta)$ as a function of θ for λ = 600 and 660 nm. SPP extraction efficiencies $\tilde{\alpha}$ and re-emission ratios $\tau(\lambda) = \frac{\Gamma_{SPP}}{\Gamma_{rad}}\tilde{\alpha}$ are given.

**Conclusion**

In this paper, we evidenced SPP-assisted visible light emission by semiconductor nanocrystals deposited on a self-assembled gold plasmonic crystal.

The self-assembled structure used offers very broadband plasmonic resonances. As a consequence, the SPP re-emission process was found to cover a spectral range of 60 nm and an angular range larger than 60° (in two lobes).

An analytical model was then proposed to describe light emission from nano-emitters coupled to plasmonic structures.

Then, an experimental method was developed to evaluate the SPP extraction efficiency and the ratio of SPP re-emission versus radiative emission. On the plasmonic crystal studied here, we found experimental SPP extraction efficiencies $\tilde{\alpha}$ of 4.5 to 6% and SPP re-emission versus radiative emission ratios $\tau$ of 15 to 20%.

The originality of the method proposed here is to use emitters as probes of the SPP surface modes. To our knowledge this is the first time grating-induced SPP extraction efficiencies are determined from photoluminescence measurements. The result characterises the grating itself and its ability to extract light.

Using fluorescent particles as probes might also enable to develop set-ups dedicated to local in-situ measurements of the SPP extraction.

**Acknowledgements**

The authors thank C'Nano Ile de France (NanoPlasmAA) and the Agence Nationale pour la Recherche (ANR Delight) for their financial support. The authors are grateful to B. Dubertret for nanocrystals synthesis, to V.M. Malasov, A. Gruzintzev, S. Chenot, L. Becerra, M. Escudier, and M. Jacquet for samples fabrication; and to D. Demaille, E. Lacaze, N. Guth, and B. Gallas for sample characterizations.